# Machine Learning Prediction of Cardiovascular Risk in Type 1 Diabetes Mellitus Using Radiomics Features from Multimodal Retinal Images

**Running head**: Radiomics-based FR, OCT, and OCTA Cardiovascular Risk Prediction

Ariadna Tohà-Dalmau[1], Josep Rosinés-Fonoll[2], Enrique Romero[1,3], Ferran Mazzanti[4], Ruben Martin-Pinardel[5], Sonia Marias-Perez[2], Carolina Bernal-Morales[2,5,6], Rafael Castro-Dominguez[2], Andrea Mendez[2], Emilio Ortega[5,6,7], Irene Vinagre[5,6,7], Marga Gimenez[5,6,7], Alfredo Vellido[1,3] and Javier Zarranz-Ventura[2,5,6,7]

[1] Department of Computer Science, Universitat Politècnica de Catalunya (UPC), Barcelona, Spain
[2] Institut Clínic d´Oftalmología (ICOF), Hospital Clínic de Barcelona, Barcelona, Spain
[3] Intelligent Data Science and Artificial Intelligence (IDEAI-UPC) Research Center, Barcelona, Spain
[4] Department of Physics, Universitat Politècnica de Catalunya (UPC), Barcelona, Spain
[5] August Pi i Sunyer Biomedical Research Institute (IDIBAPS), Barcelona, Spain
[6] Diabetes Unit, Hospital Clínic de Barcelona, Spain
[7] School of Medicine, Universitat de Barcelona, Spain

**Corresponding author & study coordinator:**
Javier Zarranz-Ventura MD PhD MSc FEBO
Hospital Clínic of Barcelona
C/ Sabino Arana 1
Barcelona, Spain 08028
zarranz@clinic.cat

**Financial support**
JZV acknowledge funding from Fundació La Marató de TV3, La Marató 2015, Diabetis i Obesitat (grant number 201633.10) and Instituto de Salud Carlos III through the projects PI18/00518 and PI21/01384 co-funded by European Union.
AV and ER acknowledge funding from Spanish research grant PID2022-143299OB-I00/AEI/10.13039/501100011033/FEDER, UE.
FM thanks the Generalitat de Catalunya for the grant Grup de Recerca SGR-Cat2021 with reference 2021SGR-01411, and by the Ministerio de Ciencia e Innovación MCIN/AEI/10.13039/501100011033 (Spain) under Grant No. PID2023-147469NB-C21

**Abbreviations and Acronyms**
ML = machine learning; CV = cardiovascular; DM = diabetes mellitus; T1DM = type 1 diabetes mellitus; T2DM = type 2 diabetes mellitus; FR = fundus retinography; OCT = optical coherence tomography; OCTA = OCT angiography; AUC = area under the curve; SD = standard deviation; CVD = cardiovascular disease; ESC = european society of cardiology; SCP = superficial capillary plexus; DCP = deep capillary plexus; FAZ = foveal avascular zone; LR = logistic regression; LDA = linear discriminant analysis; SVC = support vector classifier; rbf = radial basis function; MLP = multilayer perceptron; RF = random forest; ROC = receiver operating characteristic.

**ABSTRACT**

**Purpose:** To develop a machine learning (ML) algorithm capable of determining cardiovascular (CV) risk in multimodal retinal images from patients with type 1 diabetes mellitus (T1DM), distinguishing between moderate, high, and very high-risk levels.

**Design:** Cross-sectional analysis of a retinal image dataset from a previous prospective OCTA study (ClinicalTrials.gov NCT03422965).

**Participants:** Patients with T1DM included in the progenitor study.

**Methods:** Radiomic features were extracted from fundus retinography (FR), optical coherence tomography (OCT), and OCT angiography (OCTA) images, and ML models were trained using these features either individually or combined with clinical data (demographics and systemic data, OCT+OCTA commercial software metrics, ocular data, bloods data). Different data combinations were tested to determine the CV risk stages, defined according to international classifications.

**Main outcome measures:** Area under the receiver operating characteristic curve (AUC) mean and standard deviation (SD) for each ML model and each data combination.

**Results:** A dataset of 597 eyes (359 individuals) was analyzed. Models trained only with the radiomic features achieved AUC values of (0.79 ± 0.03) to identify moderate risk cases from high and very high cases, and (0.73 ± 0.07) for distinguishing between high and very high risk cases. The addition of clinical variables improved all AUC values, obtaining (0.99 ± 0.01) for identifying moderate cases, and (0.95 ± 0.02) for differentiating between high and very high risk cases. For very high CV risk, radiomics combined with OCT+OCTA metrics and ocular data achieved an AUC of (0.89 ± 0.02) without systemic data input. The performance of the models was similar in unilateral and bilateral eye image datasets.

**Conclusions:** Radiomic features obtained from retinal images are helpful to discriminate and classify CV risk labels, differentiating risk categories. The addition of demographics and systemic data combined with ocular data differentiate high from very high CV risk cases, and interestingly OCT+OCTA metrics with ocular data identify very high CV cases without systemic data input. These results reflect the potential of this oculomics approach for CV risk assessment.

# INTRODUCTION

Cardiovascular diseases (CVDs) are the leading cause of death globally, responsible for an estimated 17.9 million deaths in 2019.[1] CVDs are related to heart and blood vessel disorders, such as coronary artery diseases, heart attacks and strokes, and their main risk factors include insufficient physical activity, unhealthy eating patterns, smoking and other systemic diseases, such as diabetes mellitus.[2] Diabetes mellitus affects a large portion of the population, and the global prevalence of diabetes in people between 20 and 79 years old is estimated to be 10.5% in 2021, affecting 536.6 million people.[3] Type 1 DM is a chronic disease that affects approximately 2% of overall diabetes cases, but starts early in life and often leads to vascular complications.[4,5] Early detection and management of CVD risk factors are crucial in preventing progression and reducing mortality rates.

Cardiovascular risk is categorized as moderate, high, and very high, in patients with diabetes, as established and defined by the European Society of Cardiology (ESC).[6] This risk classification is calculated using clinical factors such as the duration of DM, the existence of organ damage and other major risk factors (age, hypertension, dyslipidaemia, smoking and obesity), which detail an adequate assessment of the individual status of each patient. Recently, there is a growing body of evidence that reveals strong associations between retinal changes and the development and progression of cardiovascular diseases[7], suggesting that retinal imaging could be a valid non-invasive method for the objective assessment of cardiovascular risk, in a recently created research field called oculomics.[8]

The use of radiomic features from retinal images for classification purposes is of particular interest, as it allows the extraction of relevant information from fundus retinography (FR), optical coherence tomography (OCT) and OCT angiography (OCTA) images, providing a large number of image-describing quantitative features and enabling the consideration of disease characteristics that could not be detected by other means.[9] The use of radiomics applied to OCT images has been evaluated to predict treatment response in neovascular age-related macular degeneration[10,11] and diabetic macular edema,[12,13] as well as to identify signs of intraocular inflammation.[14] We have previously applied this methodology successfully to identify diabetic retinopathy stages in retinal images.[15]

Machine learning (ML) techniques have been applied to develop clinical prediction models for cardiovascular diseases in recent years using clinical data, providing advantages over traditional statistical models such as greater model flexibility and the ability to handle larger datasets. The results obtained have generally outperformed those of non-ML models with which they were compared,[16] a fact that encourages their application in this field.

In ophthalmology, the efforts have been directed to investigate this potential to estimate cardiovascular risk in FR images,[17-21] but scarce data is available on OCT and OCTA images[22], probably related to the lack of standardized datasets. Existing studies on OCT[23] and OCTA[24,25] have focused on predicting individual cardiovascular risk factors rather than performing cardiovascular risk classification. Additionally, most research in this field does not use radiomic features, despite their potential to enhance ML-based classification models. To the best of our knowledge, no previous study has applied ML techniques to classify cardiovascular risk in T1DM patients using OCT and OCTA images.

This study aims to assess the effectiveness of using the radiomic features from FR, OCT, and OCTA images for classifying cardiovascular risk in a dataset from a large T1DM cohort.[9,26-28] ML algorithms will be applied to these radiomic features in each retinal image type, and subsequently, the addition of different combinations of clinical data will be investigated to evaluate the performance and robustness of the models with these additional data. In all cases, a feature selection process will be applied with the aim of estimating the most relevant attributes and parameters required for best performance of the models, and to identify and compare them with the items used by the ESC classification in the general population.

## METHODS

### Dataset Description

The retinal images dataset was collected as part of a prospective OCTA trial (ClinicalTrials.gov NCT03422965).[15] All participants provided signed informed consent. The baseline images of 596 patients (1175 eyes) from this prospective cohort were used as initial dataset. Exclusion criteria were applied[9], and eyes with insufficient image quality were excluded (defined as signal strength index <7 by the commercial software), with a total number of 439 patients (726 eyes). From this dataset, patients and controls in which cardiovascular risk level calculation was not possible due to individual items missing data were excluded. The final dataset included 359 patients (597 eyes) with complete information and included 36 patients with moderate risk (67 eyes), 141 patients with high risk (230 eyes), and 182 patients with very high risk (300 eyes). A consolidated standard for reporting trials (CONSORT)-style diagram is presented in Figure 1.

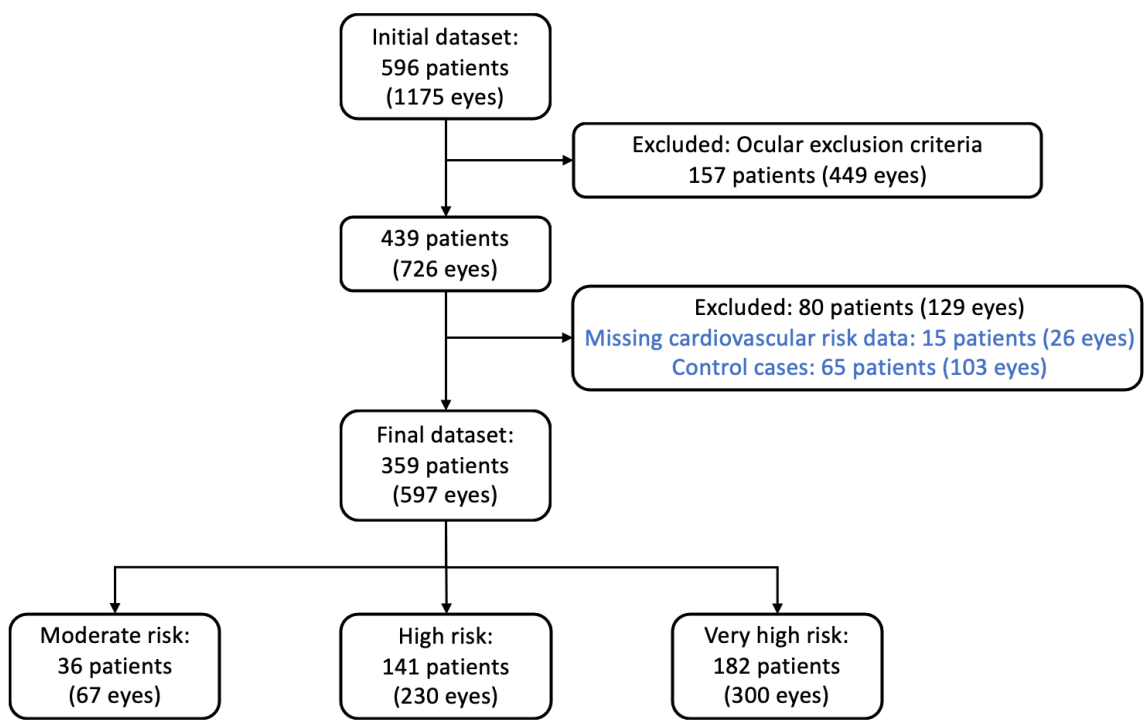

**Figure 1. Consolidated standard for outcome reporting trials (CONSORT) flowchart diagram of included and excluded patients and eyes.**

### Cardiovascular risk classifications

Individual patients were classified according to the ESC cardiovascular risk classification.[6] Three groups were created: "moderate risk", "high risk", and "very high risk" (Supplementary Table 1). Two separate classification tasks were considered: Classification task 1 ("moderate risk" vs "high" or "very high risk") and classification task 2 ("high risk" vs "very high risk"). The number of patients in each group for all the proposed classification tasks is shown in Table 1.

**Table 1: Number of patients and eyes for each group in each classification task.** The negative group represents lower cardiovascular risk, while the positive group represents higher risk.

| Classification Task | Negative | Positive |
|---|---|---|
| 1<br>Diagnosis of High and Very high CV risk | 36 (67 eyes) | 323 (530 eyes) |
| 2<br>Diagnosis of Very high CV risk | 141 (230 eyes) | 182 (300 eyes) |

**Radiomics, Imaging and Clinical Features**

Radiomic features were extracted from all retinal images in each eye: FR, OCT, OCTA 3x3 mm Superficial Capillary Plexus (SCP), OCTA 3x3 mm Deep Capillary Plexus (DCP), OCTA 6x6 mm SCP and OCTA 6x6 mm DCP (Figure 2)[7]. A total of 91 radiomic attributes were extracted from these images, including the 10$^{th}$ percentile, 90$^{th}$ percentile, energy, interquartile range, kurtosis, maximum, mean, mean absolute deviation, median, minimum, range, robust mean absolute deviation, robust mean squared, root mean squared, skewness, total energy, variance, etc.[7]

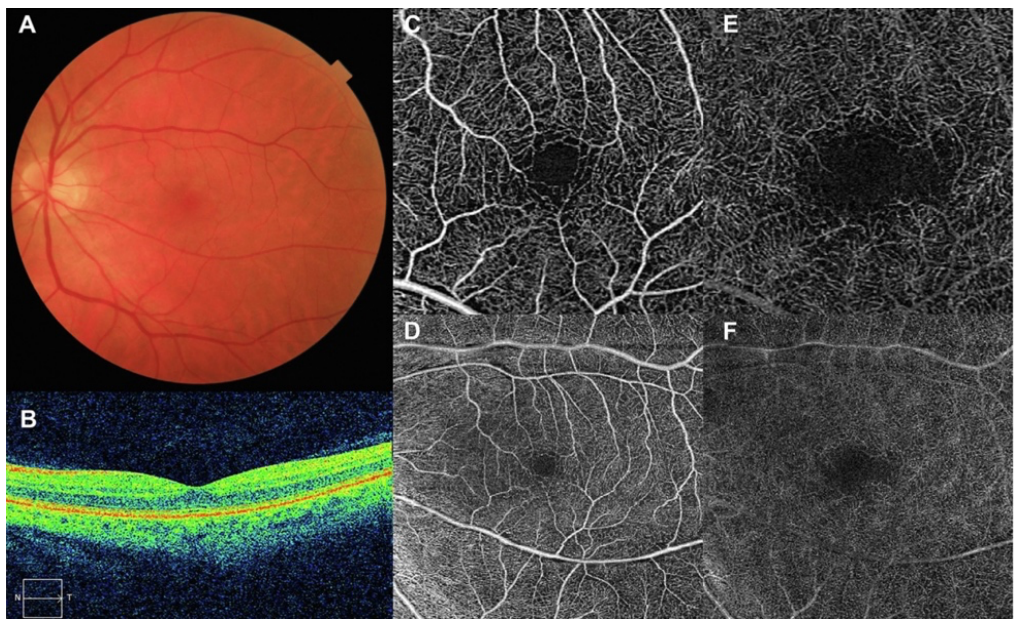

**Figure 2. Retinal images collected for each individual eye included in the study dataset.** A: Fundus retinography (FR); B: Structural optical coherence tomography (OCT) macular scan; C: OCT angiography (OCTA) 3x3 mm Superficial Capillary Plexus (SCP); D: OCTA 3x3 mm Deep Capillary Plexus (DCP); E: OCTA 6x6 mm Superficial Capillary Plexus (SCP); F: OCTA 6x6 mm Deep Capillary Plexus (DCP). (FR: Topcon DRI-Triton, Topcon Corp, Japan; OCT and OCTA: Cirrus, Carl Zeiss Meditec, Dublin, CA)(adapted from Carrera-Escale et al.).[9]

Data groups were created to categorize the additional attributes, presented in Figure 3. These groups included radiomics, commercial software OCT+OCTA metrics, ocular data, demographics and systemic data, as detailed below.

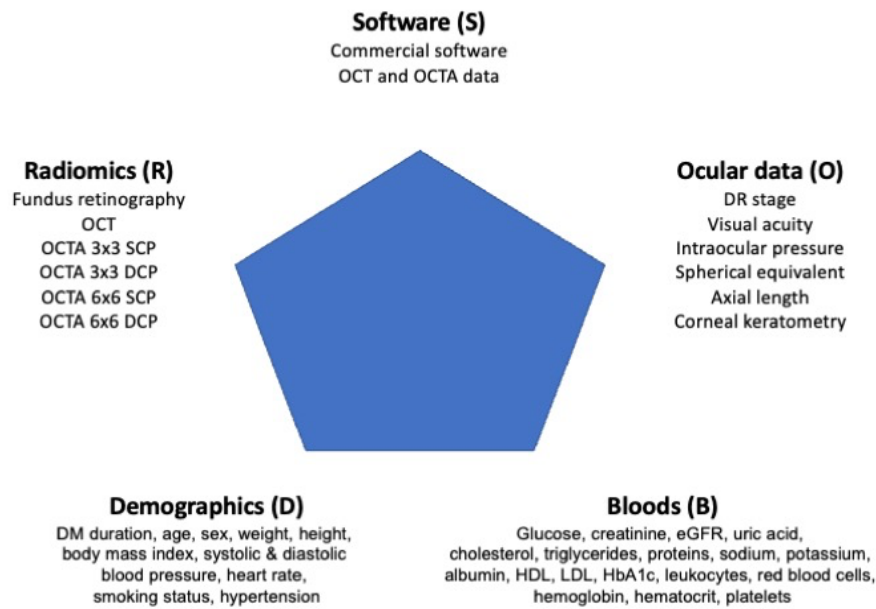

**Figure 3. Data groups scheme of the attributes used for the classification tasks.**

Standard OCT and OCTA metrics were obtained using the commercially available built-in software (Angioplex Zeiss), and included: central retinal thickness, retinal volume, average thickness in each ETDRS sector, vascular density, perfusion density, foveal avascular zone (FAZ) area, perimeter and circularity in 3x3mm and 6x6mm images.

Clinical data included demographics and systemic data (DM duration, age, sex, weight, height, body mass index -BMI-, systolic blood pressure, diastolic blood pressure, heart rate, smoking status, and arterial hypertension -HTA-), bloods data (glucose, creatinine, estimated glomerular filtration rate, uric acid, total cholesterol, triglycerides, total proteins, sodium, potassium, urinary albumin, albumin, high-density lipoprotein -HDL-, cholesterol, leukocytes, red blood cells, hemoglobin, hematocrit, platelets, mean cholesterol, mean low-density lipoprotein -LDL-, and mean glycosylated hemoglobin -HbA1c-) and ocular data (diabetic retinopathy stage, visual acuity, intraocular pressure, spherical equivalent, axial length, and corneal keratometry).

The data combinations investigated in the study were:
- R: Radiomics of retinal images.
- R+S: Radiomics of retinal images + commercial software OCT and OCTA data.
- R+O: Radiomics of retinal images + Ocular data.
- R+S+O: Radiomics of retinal images + commercial software OCT and OCTA data + Ocular data.
- R+D: Radiomics of retinal images + Demographics and systemic data.
- R+D+B: Radiomics of retinal images + Demographics and systemic data + Bloods data.
- R+D+O: Radiomics of retinal images + Demographics and systemic data + Ocular data.
- R+D+O+B: Radiomics of retinal images + Demographics and systemic data + Ocular data + Bloods data.
- ALL: Radiomics of retinal images + Demographics and systemic data + Ocular data + Bloods data + commercial software data.

Radiomics missing values were replaced by the median value of the corresponding attribute; continuous data were normalized (mean=0, standard deviation=1); and one-hot encoding was applied on categorical data.

**Machine Learning Models**
A set of standard ML and related statistical techniques were used to perform the classification. The models initially compared were Logistic Regression (LR), Linear Discriminant Analysis (LDA), Support Vector Classifiers (SVC) using linear (SVC-linear) and radial basis function (SVC-rbf) kernels, Multilayer Perceptron (MLP), and Random Forest (RF). Those that provided significantly lower AUC values at the first stages of model development were discarded.

The proposed ML models were evaluated with a double *k-m*-fold cross validation technique, where the data was first divided into k=5 partitions, and each partition was used once as the test set, while the remaining data was used for model selection. For each iteration, the training data was further split into m=4 subsets, which were used for hyperparameter optimization and feature selection by training the model on 3 of the subsets and validating it on the remaining one. The outer *k*-fold validation evaluates the model's overall performance by testing it on data that was held out during both training and model adjustment. This technique allows for the use of all the data for training, validation, and testing the model. This approach is particularly useful given the relatively limited number of patients available. To reduce the risk of bilaterality bias, both eyes of the same patient were always assigned to the same partition in the double *k-m*-fold cross validation technique to ensure that model performance was assessed on independent data.

**Hyperparameters and Feature Selection, Model Optimization and Performance**
The procedure for all classification tasks, data combinations, and proposed ML models is presented in Figure 4.

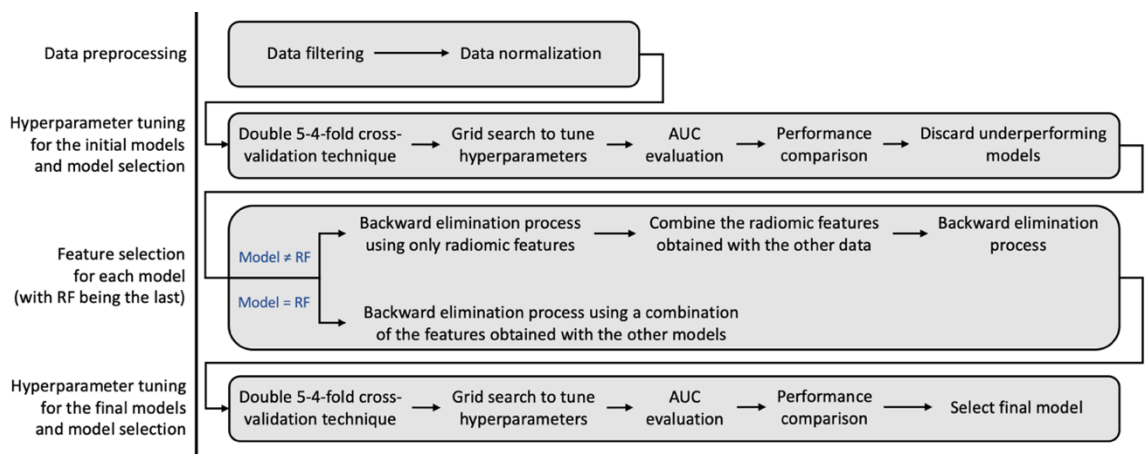

**Figure 4. Pipeline of the methodology followed to train and optimize the machine learning (ML) models.** This process is repeated for all the proposed classification tasks, data combinations and models. AUC = Area Under the ROC Curve; RF = Random Forest.

After data preprocessing, an initial tuning of all the proposed models was performed, selecting the hyperparameters leading to the highest AUC values in each case. By comparing the results of the different models, it was possible to discard those with significantly worse performance (LDA and MLP) for both classification tasks. The resulting configuration was used to carry out feature selection through a backward elimination process, divided in two distinct steps. First, only the radiomic data was used, selecting the attributes that are relevant for each problem without adding any additional information (i.e. using the data combination R). Second, once the relevant radiomic features for each problem and model were selected, feature selection was performed for all data combinations. For the RF model, a combination of the attributes that were found to be relevant for the other models was used. The backward elimination process was repeated for each data combination, classification task, and model, resulting in a final set of attributes that typically ranged between 5 and 20.

After determining the features that will be used for the classification, hyperparameters selection was repeated to optimize the models. Finally, model performance was evaluated by calculating the mean AUC values and by constructing the mean receiver operating characteristic (ROC) curves.

**Statistical tests**
To compare the different ML models, pairwise DeLong tests were performed for each classification task, using the data combinations R, R+S+O, and all the data, as these are of greatest interest for the use of radiomics. These tests were also used to compare the results from different data combinations in pairs to determine which data groups contributed most significantly to the classification; this comparison was conducted only with the SVC-rbf model, which was selected since it was the model that, overall, obtained best results.

**Feature importance**
To determine which features are most relevant to perform the classifications, two different approaches are used. On one hand, the models of greatest interest for the study are trained again, but this time using only one type of image data, either FR, OCT, or OCTA, along with the necessary clinical data in each case. Additionally, the different OCTA images available (3x3 and 6x6, deep and superficial in both cases) are also tested separately. The AUC obtained for each combination is compared to demonstrate the advantage of using OCT or OCTA for diagnosis instead of retinal fundus images. On the other hand, the SHapley Additive exPlanation (SHAP) technique is applied to the final models to assess the importance of each feature considered necessary after the backward elimination process.

## RESULTS

The performance of the final ML models was assessed based on the AUC values. This section provides a detailed comparison of model effectiveness across both classification tasks, evaluating different data combinations and supporting the analysis with statistical tests and feature importance assessment.

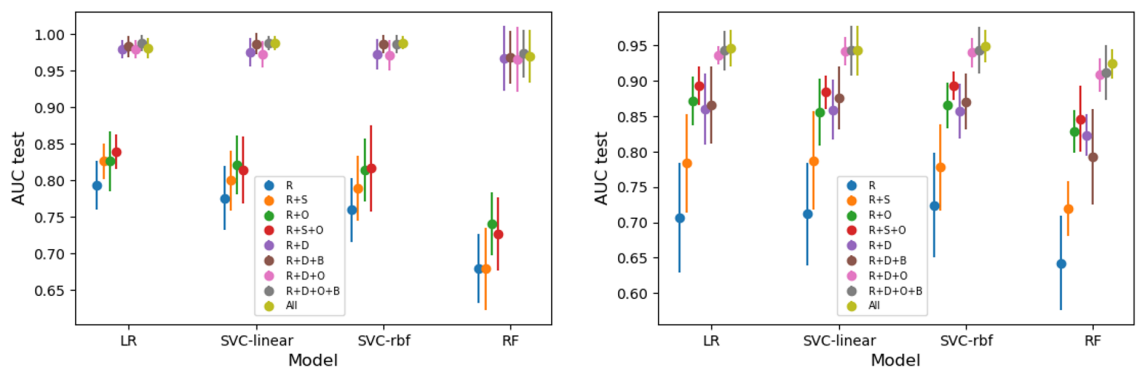

**Figure 5. Machine learning (ML) models performance for all the classification tasks and data combinations, representing area under the curve (AUC) values.** Left: AUC values for distinguishing between moderate risk and high or very high risk. Right: AUC values for distinguishing between high risk and very high risk. All these results are extracted using both eyes per patient. R = Radiomics of retinal images; S = Commercial Software OCT and OCTA data; O = Ocular data; D = Demographics and systemic data; B = Blood analysis data.

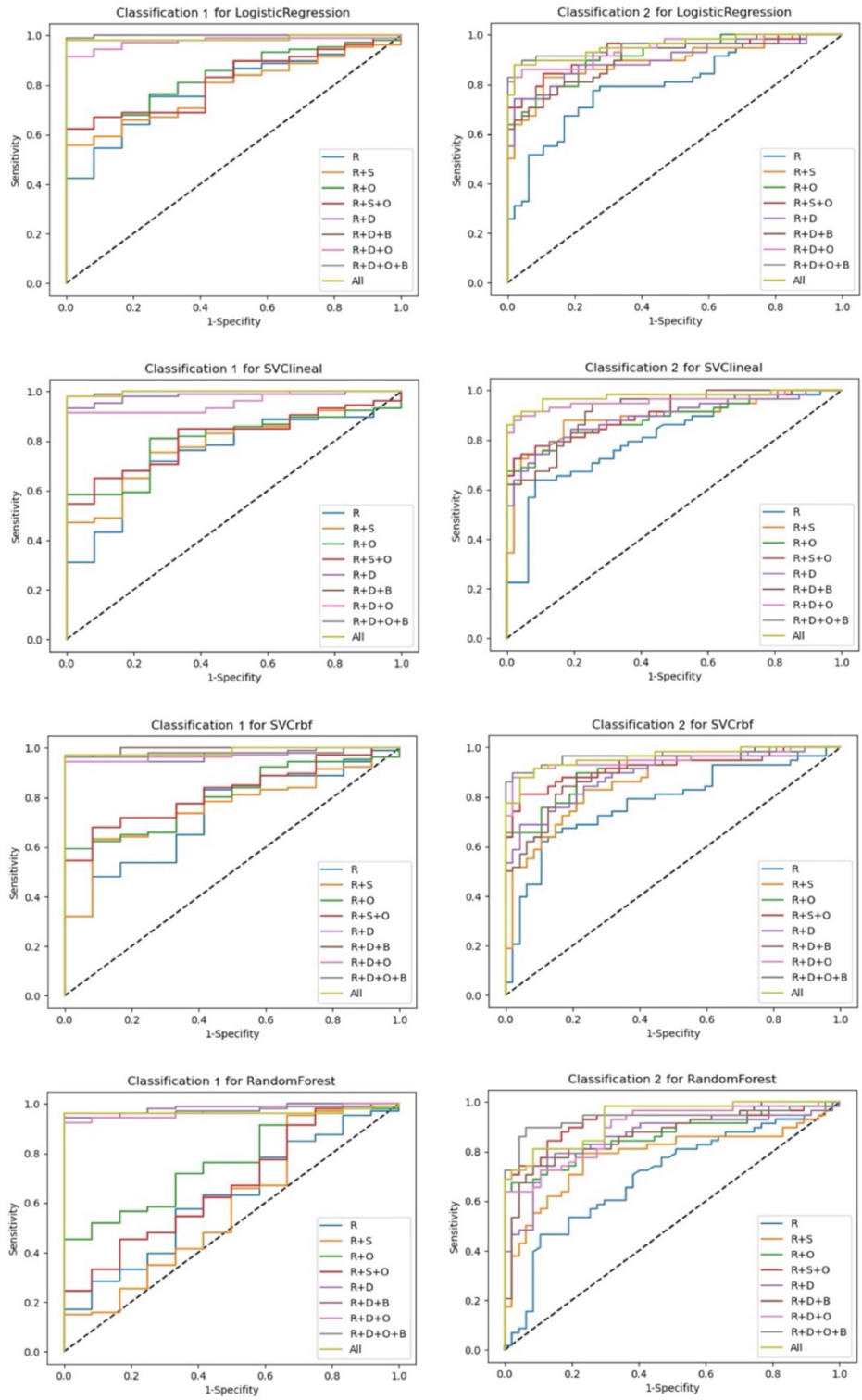

**Figure 6. Receiver operating characteristic (ROC) curves of models' performance for all the classification tasks and data combinations.** Left column: ROC curves discerning between moderate risk and high or very high risk. Right column: ROC curves for distinguishing between high risk and very high risk. Top row: Logistic Regression (LR) model. Second row: Support Vector Classifier using linear kernel (SVC-linear) model. Third row: Support Vector Classifier using radial basis function kernel (SVC-rbf) model. Bottom row: Random Forest (RF) model. All these results are extracted using both eyes per patient. R = Radiomics of retinal images; S = Data from the commercial software; O = Ocular data; D = Demographic and systemic data; B = Blood analysis data.

**Table 2: Models performance for all classification tasks and data combinations, including a comparison between using 1 or 2 eyes per patient.** The best results for each problem are highlighted.

| Class. Task | Data | LR | | SVC-Linear | | SVC-rbf | | RF | |
|---|---|---|---|---|---|---|---|---|---|
| | | 1 Eye | 2 Eyes | 1 Eye | 2 Eyes | 1 Eye | 2 Eyes | 1 Eye | 2 Eyes |
| **1 Diagnosis of High and Very high CV risk** | R | 0.75 ± 0.04 | **0.79 ± 0.03** | 0.79 ± 0.06 | 0.78 ± 0.04 | 0.76 ± 0.04 | 0.76 ± 0.04 | 0.72 ± 0.07 | 0.68 ± 0.05 |
| | R+S | 0.83 ± 0.05 | **0.83 ± 0.02** | 0.82 ± 0.04 | 0.80 ± 0.04 | 0.82 ± 0.06 | 0.79 ± 0.04 | 0.70 ± 0.10 | 0.68 ± 0.06 |
| | R+O | 0.82 ± 0.07 | 0.83 ± 0.04 | 0.78 ± 0.03 | 0.82 ± 0.04 | **0.84 ± 0.04** | 0.81 ± 0.04 | 0.72 ± 0.05 | 0.74 ± 0.04 |
| | R+S+O | 0.84 ± 0.03 | **0.84 ± 0.02** | 0.82 ± 0.08 | 0.81 ± 0.05 | 0.83 ± 0.05 | 0.82 ± 0.06 | 0.76 ± 0.06 | 0.73 ± 0.05 |
| | R+D | **0.98 ± 0.01** | **0.98 ± 0.01** | 0.98 ± 0.01 | 0.97 ± 0.02 | 0.97 ± 0.03 | 0.97 ± 0.02 | 0.96 ± 0.04 | 0.97 ± 0.04 |
| | R+D+B | 0.98 ± 0.01 | 0.98 ± 0.01 | 0.98 ± 0.01 | **0.99 ± 0.01** | 0.98 ± 0.01 | **0.99 ± 0.01** | 0.97 ± 0.04 | 0.97 ± 0.04 |
| | R+D+O | 0.97 ± 0.01 | **0.98 ± 0.01** | 0.98 ± 0.02 | 0.97 ± 0.02 | 0.97 ± 0.02 | 0.97 ± 0.02 | 0.98 ± 0.02 | 0.96 ± 0.04 |
| | R+D+B+O | 0.98 ± 0.01 | **0.99 ± 0.01** | **0.99 ± 0.01** | **0.99 ± 0.01** | **0.99 ± 0.01** | **0.99 ± 0.01** | 0.97 ± 0.04 | 0.97 ± 0.03 |
| | All | 0.98 ± 0.01 | 0.98 ± 0.01 | **0.99 ± 0.01** | **0.99 ± 0.01** | **0.99 ± 0.01** | **0.99 ± 0.01** | 0.97 ± 0.04 | 0.97 ± 0.04 |
| **2 Diagnosis of Very high CV risk** | R | 0.70 ± 0.08 | 0.71 ± 0.08 | 0.68 ± 0.09 | 0.71 ± 0.07 | 0.73 ± 0.10 | **0.73 ± 0.07** | 0.66 ± 0.07 | 0.64 ± 0.07 |
| | R+S | 0.78 ± 0.07 | 0.78 ± 0.07 | 0.76 ± 0.09 | **0.79 ± 0.07** | 0.73 ± 0.09 | 0.78 ± 0.06 | 0.71 ± 0.06 | 0.72 ± 0.04 |
| | R+O | 0.85 ± 0.04 | **0.87 ± 0.03** | 0.85 ± 0.04 | 0.86 ± 0.05 | 0.86 ± 0.04 | **0.87 ± 0.03** | 0.86 ± 0.04 | 0.83 ± 0.03 |
| | R+S+O | 0.88 ± 0.04 | 0.89 ± 0.03 | 0.86 ± 0.01 | 0.88 ± 0.02 | 0.88 ± 0.02 | **0.89 ± 0.02** | 0.84 ± 0.04 | 0.85 ± 0.05 |
| | R+D | 0.83 ± 0.06 | 0.86 ± 0.05 | 0.83 ± 0.06 | **0.86 ± 0.04** | 0.84 ± 0.06 | **0.86 ± 0.04** | 0.79 ± 0.05 | 0.82 ± 0.03 |
| | R+D+B | 0.84 ± 0.06 | 0.87 ± 0.05 | 0.85 ± 0.06 | **0.88 ± 0.05** | 0.84 ± 0.04 | 0.87 ± 0.04 | 0.79 ± 0.07 | 0.79 ± 0.07 |
| | R+D+O | **0.94 ± 0.01** | **0.94 ± 0.01** | 0.94 ± 0.03 | 0.94 ± 0.02 | 0.92 ± 0.02 | 0.94 ± 0.02 | 0.91 ± 0.02 | 0.91 ± 0.02 |
| | R+D+B+O | **0.94 ± 0.03** | **0.94 ± 0.03** | 0.94 ± 0.04 | 0.94 ± 0.04 | 0.94 ± 0.04 | **0.94 ± 0.03** | 0.89 ± 0.04 | 0.91 ± 0.04 |
| | All | 0.94 ± 0.03 | 0.95 ± 0.03 | 0.93 ± 0.04 | 0.94 ± 0.04 | 0.94 ± 0.02 | **0.95 ± 0.02** | 0.91 ± 0.03 | 0.92 ± 0.02 |

**Classification Task 1: Detection of high and very high cardiovascular risk**

The AUC values obtained for each ML model are detailed in Table 2 and illustrated graphically in Figure 5, with their ROC curves shown in Figure 6. To distinguish between patients with moderate risk and those with high or very high risk, the highest AUC performance using only radiomic features was obtained from the LR model (AUC 0.79 ± 0.03). The addition of ocular and commercial software data improved the performance within the same model, also achieving the best results (AUC 0.84 ± 0.02). The highest performance was achieved when radiomic features, demographics, systemics, and blood analysis data were combined, using the SVC-Linear and SVC-rbf models (AUC 0.99 ± 0.01).

**Table 3: Statistical comparison of model performance.** P values from the DeLong test for both classifications using the most relevant data combinations for this study. All results are derived from models obtained using both eyes per patient. The p values that are lower than 0.05 are highlighted. R = Radiomics of eye images; S = Data from the commercial software; O = Ocular data.

| Class. Task | Data | P Value (DeLong test) | | | | | |
|---|---|---|---|---|---|---|---|
| | | LR-SVClinear | LR-SVCrbf | LR-RF | SVClinear-SVCrbf | SVClinear-RF | SVCrbf-RF |
| 1 Diagnosis of High and Very high CV risk | R | 0.431 | 0.125 | **0.000** | 0.507 | **0.000** | **0.003** |
| | R+S+O | 0.240 | 0.370 | **0.000** | 0.957 | **0.001** | **0.004** |
| | All | 0.276 | 0.238 | 0.506 | 0.950 | 0.238 | 0.225 |
| 2 Diagnosis of Very high CV risk | R | 0.896 | 0.682 | 0.130 | 0.772 | 0.087 | **0.046** |
| | R+S+O | 0.511 | 0.991 | **0.014** | 0.445 | **0.035** | **0.008** |
| | All | 0.839 | 0.837 | 0.084 | 0.704 | 0.233 | **0.037** |

**Table 4: Statistical comparison of model performance for different data combinations.** P values from the DeLong test for both classification tasks, with classification 1 in blue (diagnosis of high and very high risk) and classification task 2 in orange (diagnosis of very high risk), using the SVC-rbf model. All results are derived from models obtained using both eyes per patient. The p values that are lower than 0.05 are highlighted. R = Radiomics of retinal images; S = Data from the commercial software; O = Ocular data; D = Demographic and systemic data; B = Blood analysis data.

| | R | R+S | R+O | R+S+O | R+D | R+D+B | R+D+O | R+D+B+O | All |
|---|---|---|---|---|---|---|---|---|---|
| R | | 0.232 | **0.025** | 0.052 | **0.000** | **0.000** | **0.000** | **0.000** | **0.000** |
| R+S | 0.157 | | 0.303 | 0.354 | **0.000** | **0.000** | **0.000** | **0.000** | **0.000** |
| R+O | **0.000** | **0.002** | | 0.949 | **0.000** | **0.000** | **0.000** | **0.000** | **0.000** |
| R+S+O | **0.000** | **0.000** | 0.063 | | **0.000** | **0.000** | **0.000** | **0.000** | **0.000** |
| R+D | **0.000** | **0.006** | 0.701 | **0.040** | | 0.138 | 0.928 | 0.138 | 0.074 |
| R+D+B | **0.000** | **0.001** | 0.797 | 0.196 | 0.558 | | 0.121 | 0.987 | 0.793 |
| R+D+O | **0.000** | **0.000** | **0.000** | **0.000** | **0.000** | **0.000** | | 0.121 | 0.065 |
| R+D+B+O | **0.000** | **0.000** | **0.000** | **0.001** | **0.000** | **0.000** | 0.789 | | 0.774 |
| All | **0.000** | **0.000** | **0.000** | **0.000** | **0.000** | **0.000** | 0.452 | 0.751 | |

The DeLong test analysis shows that the results obtained by the ML models using all data were similar, while in the R and R+S+O cases the RF model presented a different behaviour than the rest, as shown in Table 3. Regarding the differences among data groups, two clusters were observed: {R, R+S, R+O, R+S+O} and {R+D, R+D+B, R+D+O, R+D+B+O, All}, as reported in Table 4. These two clusters are characterized by the presence or absence of demographic and systemic data. Although the pairwise test for R vs R+O yielded a p value below 0.05, indicating some difference, it was notably higher than most other comparisons.

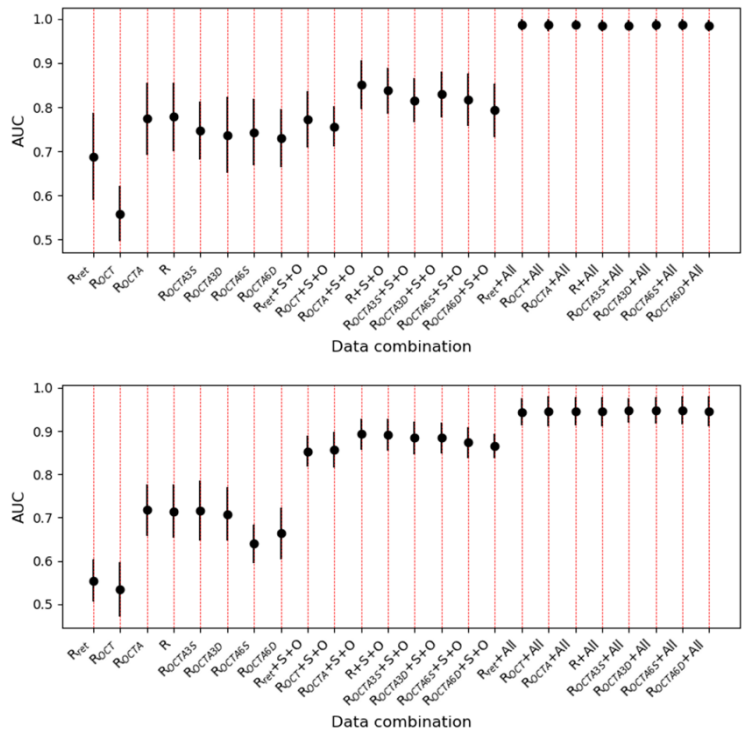

**Figure 7. Machine learning (ML) models performance for both classification tasks, evaluated using the best data combinations but changing the type of images used in each case.** It presents area under the curve (AUC) values. Top: AUC values for distinguishing between moderate risk and high or very high risk. Bottom: AUC values for distinguishing between high risk and very high risk. All these results are extracted using both eyes per patient. R = Radiomics of retinal images; S = Commercial Software OCT and OCTA data; O = Ocular data; D = Demographics and systemic data; B = Blood analysis data.

The AUC values used to compare the relevance of the different types of images are shown in Figure 7, where the radiomic features included in the data combinations R, R+S+O and All are used separately. The difference is most noticeable when using only the radiomic features, with a clearly better performance of the models trained on OCTA images, without significant variations among the different OCTA scan sizes and plexus analyzed. When using the R+S+O data group, OCTA images maintain the highest performance of the models, with minor differences with other imaging techniques. Finally, when all the data is included in the models, the results are equivalent regardless of the retinal image type.

The attributes that achieved the best performance are shown in Table 5, along with their mean and standard deviation in the class of interest. These features are also enumerated below, alongside their SHAP values in parentheses, which allow us to rank them from most to least relevant. When including all the available data, the final attributes are DM duration (SHAP value: 0.088), non-smoker (0.053), total cholesterol (0.034), retinopathy with degree 1 (0.030), OCTA 3x3 deep skewness (0.016), BMI (0.013), height (0.013), OCT interquartile range (0.012), triglycerides (0.006), external upper sector thickness (SE, extracted from the OCT image, 0.003) and mean HbA1c (0.002). With the data combination R+S+O the final attributes are OCTA 3x3 superficial kurtosis (0.156), retinopathy with degree 1 (0.148), vascular density 3x3 (VC 3x3, extracted from the OCTA image, 0.118), central thickness of the ETDRS grid (0.080), OCTA 3x3 superficial variance (0.057), OCTA 3x3 deep skewness (0.050), OCTA 6x6 deep skewness (0.041) and internal nasal sector thickness (0.032).

**Table 5: Statistical data for each feature in the different classifications, considering only those necessary for training the final model.** The mean and standard deviation of the features are displayed. Categorical variables take the value 1 for affirmative and 0 for negative, and radiomic data have no units.

|  |  | Classification Task 1 | | Classification Task 2 | |
|---|---|---|---|---|---|
| **Feature** | **Data group** | Moderate risk | High and very high risk | High risk | Very high risk |
| Duration DM (years) | Demographics and systemics | 5.65 ± 2.65 | 21.32 ± 10.07 | 15.36 ± 8.18 | 25.97 ± 8.90 |
| No smoker | Demographics and systemics | 1.00 ± 0.00 | 0.60 ± 0.49 | - | - |
| Total cholesterol (mg/dL) | Blood analysis | 154.97 ± 21.68 | 179.23 ± 29.31 | 177.79 ± 28.73 | 180.41 ± 29.77 |
| Height (m) | Demographics and systemics | 1.74 ± 0.10 | 1.70 ± 0.09 | - | - |
| BMI (kg/m$^3$) | Demographics and systemics | 23.05 ± 2.54 | 24.87 ± 3.72 | - | - |
| OCT interquartile range | Radiomics | -0.22 ± 0.22 | -0.14 ± 0.32 | - | - |
| Mean HbA1c (%) | Blood analysis | 7.31 ± 0.98 | 7.82 ± 1.96 | - | - |
| No Diabetic Retinopathy | Oculars | 1.00 ± 0.00 | 0.61 ± 0.49 | 1.00 ± 0.00 | 0.31 ± 0.47 |
| OCTA 3x3 deep skewness | Radiomics | -0.45 ± 0.23 | -0.22 ± 0.31 | - | - |
| SE (μm) | Commercial software | 282.71 ± 14.53 | 283.87 ± 22.19 | - | - |
| Triglycerides (mg/dL) | Blood analysis | 59.78 ± 19.36 | 83.20 ± 52.80 | 74.55 ± 31.39 | 90.24 ± 64.44 |
| HTA | Demographics and systemics | - | - | 0.03 ± 0.17 | 0.15 ± 0.36 |
| OCTA 6x6 superficial kurtosis | Radiomics | - | - | -0.38 ± 0.29 | -0.38 ± 0.28 |
| Sodium (mEq/L) | Blood analysis | - | - | 140.69 ± 2.13 | 140.12 ± 2.11 |
| SI (μm) | Oculars | - | - | 327.68 ± 16.80 | 324.79 ± 16.24 |
| Ex-smoker | Demographics and systemics | - | - | 0.12 ± 0.33 | 0.21 ± 0.41 |
| P 3X3 (mm) | Oculars | - | - | 1.99 ± 0.50 | 2.14 ± 0.54 |
| HDL cholesterol (mg/dL) | Blood analysis | - | - | 62.32 ± 16.54 | 59.24 ± 18.70 |
| Glucose (mg/dL) | Blood analysis | - | - | 153.19 ± 69.89 | 164.92 ± 76.25 |

**Classification Task 2: Detection of very high cardiovascular risk**
The AUC values obtained for each ML model are detailed in Table 2 and illustrated graphically in Figure 5, with their ROC curves shown in Figure 6. To distinguish between patients with high risk and those with very high risk, the highest AUC performance using only radiomic features was obtained by the SVC-rbf model (AUC $0.73 \pm 0.07$). The incorporation of ocular and commercial software data improved this performance, with the same model achieving the best results (AUC $0.89 \pm 0.02$) without input of demographics or systemic data. Again, the highest performance was obtained with the addition of all the clinical data, using the SVC-rbf model again (AUC $0.95 \pm 0.02$).

The DeLong test analysis shows that the results obtained by the ML models using only radiomics and all data were similar, except for the pairwise comparison between SVC rbf and RF, which yielded p values slightly below 0.05. In the R+S+O case, the RF model presented a different behaviour than the other combinations of data, as shown in Table 3. Regarding the differences among data groups, three clusters of statistically equivalent results were observed: {R, R+S}, {R+O, R+S+O, R+D, R+D+B} and {R+D+O, R+D+B+O, All}. The first group included only radiomics or commercial software data. The second one included demographics and systemic or ocular data, but not the combination of both. Within this group, the comparison between R+D and R+S+O resulted in a p value slightly below 0.05, though they can still be considered similar given the relative magnitude of differences observed in other comparisons. The third group contained both demographic and systemic and ocular data.

The AUC values used to compare the relevance of the different types of images are shown in Figure 7. As in the other classification, the difference is most noticeable when using only the radiomic features, with a clearly better performance of the models trained on OCTA images. In this case, the 3×3 images stand out compared to the 6×6 images. When using the R+S+O data group, OCTA images also stand out, but the difference is less pronounced. Finally, when using all the data, the results are equivalent regardless of the images used.

The attributes that achieved the best performance are shown in Table 5, along with their mean and standard deviation in the class of interest. These features are also enumerated below, alongside their SHAP values in parentheses, which allow us to rank them from most to least relevant. When including all the available data, the final attributes are diabetic retinopathy with degree 1 (0.257), DM duration (0.104), HTA (0.044), ex-smoker (0.029), HDL cholesterol (0.025), total cholesterol (0.020), ZAF perimeter 3x3 (P 3x3, extracted from the OCTA image, 0.019), sodium (0.018), triglycerides (0.014), internal upper sector thickness (SI, extracted from the OCT image, 0.013), OCTA 6x6 superficial kurtosis (0.011) and glucose (0.003). With the data combination R+S+O the final attributes are diabetic retinopathy with degree 1 (0.225), diabetic retinopathy with degree 2 (0.060), vascular density 3x3 (VC 3x3, extracted from the OCTA image, 0.045), OCTA 3x3 superficial robust mean absolute deviation (0.018), OCTA 6x6 superficial skewness (0.010), intraocular pressure (0.008), perfusion density 3x3 (PC 3x3, extracted from the OCTA image, 0.007), OCT kurtosis (0.007), OCTA 3x3 deep robust mean absolute deviation (0.006), OCT 90th percentile (0.003) and OCTA 6x6 superficial energy (0.002).

**Influence of Unilateral vs Bilateral Image Datasets in Models Performance**
For all classification tasks, data combinations, and models, the AUC values obtained using data from both eyes and from one randomly selected eye were very similar, with a slightly lower performance when using only one eye. Additionally, the standard deviation of the AUCs was higher when working with data from a single eye, as shown in Table 2.

**DISCUSSION**

The results of this study highlight the effectiveness of ML applied to radiomic features, extracted from FR, OCT, and OCTA images, in combination with clinical data to predict CV risk in patients with T1DM. The performance of the proposed models ranges from good to excellent, improving when additional ocular and systemic data groups are included alongside the radiomic attributes. The data presented in this study support the application of this technology for the identification of CV risk cases in type 1 DM patients in a non-invasive fashion, based in widely available retinal image techniques.

The use of the radiomic features obtained from retinal images and ML has proven to be effective in the identification of CV risk labels. Using only the radiomics of retinal images with no additional data the performance of the models has been good enough to discriminate high and very high CV risk cases (task 1, AUCs ranging from 0.79 to 0.68) and very high CV risk cases (task 2, AUCs ranging from 0.73 to 0.66). These results align with previous studies on CV risk classification for patients with DM,[29,30] where the clinical attributes used differ from those of the ESC classification. This study is the first to specifically leverage radiomic features extracted from FR, OCT and OCTA images for this purpose, and achieving comparable results is highly promising. These results suggest that this methodology provides valuable information not available through conventional clinical assessments using only the retinal images, allowing the identification of the CV risk of these patients in a non-invasive way. This method could be used in a general setting in the community if confirmed and validated in future studies.

Focusing in the translational relevance of our findings, we followed a stepwise strategy to investigate whether is possible to improve the performance of the models with the addition of different data groups, guided by a potential clinical application. First, we aimed to determine the maximum achievable AUC when utilizing only ophthalmological data, with the combination of radiomics and data obtained in a regular ophthalmological examination. As expected, adding the commercial software OCT+OCTA metrics and ocular examination data improved the performance of the models, without the need to include demographic and systemic data. Second, we investigated the influence of the addition of demographics and systemic data, which could be collected in a medical visit or comprehensive questionnaire to evaluate the potential influence of documenting this information. Again, when demographics and systemic data were included the model's performance improved significantly across all classification tasks. And third, the combination of all available data was evaluated to determine the maximum potential of this approach, for achieving optimal results in all models. Recent advances in machine learning methodologies, such as the *TabPFN* model proposed by Hollmann et al[31] provide alternative approaches for tabular data analysis that could be applied to these classification tasks in small datasets. The evaluation of this interesting strategy falls beyond the scope of this study, but appears as a promising approach that merits further research in the future.

With regards to the classification tasks, for the diagnosis of high and very high CV risk cases together the addition of systemic data (demographics, bloods) was required in most of the models, but for isolated very high CV risk cases an adequate performance was achieved with just ocular and retinal imaging data (software, ocular data) without the addition of systemic data. This is one of the main findings of the study, as it highlights that a complete ocular examination could be an adequate strategy to identify very high CV risk cases, maybe in specific subgroups of type 1 DM patients (i.e. preoperative assessments prior to major procedures, etc.). As expected, the highest improvement in all the models was observed when demographics, ocular and systemic data were included in the models, reflecting the maximum potential of this methodology.

To provide valuable insights into the key factors related to cardiovascular risk in T1DM patients and benchmark our results with the community, a detailed analysis of the features used to train the final models is provided in Table 5. Many of the top attributes include clinical biomarkers such as HbA1c and cholesterol, which are known indicators of metabolic health and cardiovascular risk, and others included in the ESC classification. For both classifications, the most important attributes included DM duration, smoking status and the presence of diabetic retinopathy, proving the robustness of the models. In addition, we found that some individual radiomic features are strong predictors, which supports their use as a powerful tool in cardiovascular risk prediction. For example, to distinguish moderate from high and very high CV risk, the strongest predictors are the radiomic features 'OCTA 3x3 deep skewness' and 'OCT interquartile range', which provide information about the variability and distribution of retinal thickness and retinal blood flow, respectively. With regards to the latter, to differentiate between high and very high CV risk the radiomic feature 'OCTA 6x6 superficial kurtosis', which indicates the sharpness of the intensity distribution in the images, was the strongest predictor, although its importance was relatively low compared to the clinical attributes used.

Focusing on the radiomic features extracted from different types of images, this study has established that the most relevant ones are those from OCTA images, with a clearly better performance of the models using these compared to retinal fundus or OCT images. However, this difference becomes less pronounced when adding all the clinical data, reinforcing the previously stated importance of these variables. When distinguishing high from very high risk, a superior performance of the models were observed with the OCTA 3x3 images compared to OCTA 6x6 images. The relevance of this type of data has been further supported by the final attributes obtained for the R+S+O combination, which were generally extracted from OCTA 3x3 images.

In terms of the best performing ML model, we observed similar results with LR, SVC-linear, and SVC-rbf models, but worse performance with the RF model. Based on the statistical tests, we concluded that there were clusters of models in the classifications conducted as presented in the results section. From the practical point of view, we would recommend the use of the LR model, which is computationally the simplest model, if no differences between models are confirmed in future studies on this topic. With regards to the influence of using one or both eyes data, we have observed slightly higher AUC values when using bilateral imaging data and increased variability when using unilateral data, with minimal differences in the overall performance between both options as presented in Table 2.

There is limited data available on the topic of radiomics and CV risk, to benchmark our findings. The use of radiomic features from OCT images has been applied in other fields, such as the prediction of anti-vascular endothelial growth factor (VEGF) therapy responses in neovascular AMD,[10,11] and macular edema related to vascular diseases.[13] Interestingly, in these studies the performance of the models also improved with the addition of clinical data (AUC values from 0.8 to 0.95), consistently with our results. About CV risk, recent reports have highlighted the application of artificial intelligence for diagnosing CVDs through retinal imaging-based oculomics [22] (again with AUC values ranging from 0.71 to 0.87), or deep learning to assess CV risk from retinal images, finding a clear correlation between indirect parameters such as retinal-vessel calibre and other risk factors.[19] Most of these DL studies have been conducted using FR images to predict cardiovascular risk factors with good performance[17], improving their results with the addition of clinical data (AUC values ranging from 0.65 to 0.77)[21]. When comparing these reports with the current study, our results suggest that the use of radiomics derived from OCT and OCTA images enhances ML models performance.

This study presents several strengths and limitations. First, the dataset includes type 1 DM patients, whereas most of prior research either focuses on T2DM patients or does not distinguish between both types. Consequently, these findings may not be directly applicable to T2DM patients, but this specificity strengthens the study's internal validity by leveraging one of the largest reported dataset of multimodal retinal images for T1DM patients, with previous reports that describe existing associations between OCTA-derived metrics and diabetic retinopathy,[26] diabetic kidney disease,[27] and glycated hemoglobin levels.[28] Finally, although many clinical variables used in the final models are demographic, obtaining these data may require information beyond retinal imaging alone, potentially limiting the practical application of this classification method.

In conclusion, this study demonstrates that radiomic features extracted from retinal images can be processed with ML techniques to predict cardiovascular risk levels in patients with T1DM. The performance of the models improves with additional clinical data, leading to a better classification in cases of high and very high CV risk. The results presented in this study support the application of this technology as a non-invasive method for the assessment of the CV risk status in DM patients, being based in a widely available set of retinal image techniques, which could potentially be deployed at a large scale in a community setting. In this regard, the development of a web-based calculator for personalized cardiovascular disease risk prediction could be a valuable next step for a direct translational application of the results of this study.

**REFERENCES:**

1. World Health Organization. Cardiovascular diseases (cvds), 2021. Accessed: 2024-05-16.
2. Irfan H. Obesity, cardiovascular disease, and the promising role of semaglutide:Insights from the select trial. Current Problems in Cardiology 2024;49:102060.
3. Sun H, Saeedi P, Karuranga S, et al. IDF diabetes atlas: Global, regional and country-level diabetes prevalence estimates for 2021 and projections for 2045. Diabetes Research and Clinical Practice 2021;183:109119.
4. Chalakova T, Yotov Y, Tzotchev K, et al. Type 1 diabetes mellitus - risk factor for cardiovascular disease morbidity and mortality. Current Diabetes Reviews 2021;17:37–54.
5. Green A, Hede SM, Patterson CC, et al. Type 1 diabetes in 2017: global estimates of incident and prevalent cases in children and adults. Diabetologia 2021;64:2741–2750.
6. Cosentino F, Grant PJ, Aboyans V, et al. 2019 ESC guidelines on diabetes, pre-diabetes, and cardiovascular diseases developed in collaboration with the easd. European Heart Journal 2020;41:255–323.
7. Runsewe OI, Srivastava SK, Sharma S, et al. Optical coherence tomography angiography in cardiovascular disease. Progress in Cardiovascular Diseases 2024.
8. Wagner SK, Fu DJ, Faes L, et al. Insights into Systemic Disease through Retinal Imaging-Based Oculomics. Translational Vision Science & Technology 2020;9(2).
9. Carrera-Escalé L, Benali A, Rathert AC, et al. Radiomics-based assessment of OCT angiography images for diabetic retinopathy diagnosis. Ophthalmology Science 2023;3(2):100259.
10. Williamson RC, Selvam A, Sant V, et al. Radiomics-Based Prediction of Anti-VEGF Treatment Response in Neovascular Age-Related Macular Degeneration With Pigment Epithelial Detachment. Translational Vision Science and Technology 2023;12(10).
11. Kar SS, Cetin H, Srivastava SK, et al. Texture-Based Radiomic SD-OCT Features Associated With Response to Anti-VEGF Therapy in a Phase III Neovascular AMD Clinical Trial. Translational Vision Science and Technology 2024;13(1).
12. Kar SS, Sevgi DD, Dong V, et al. Multi-Compartment Spatially-Derived Radiomics from Optical Coherence Tomography Predict Anti-VEGF Treatment Durability in Macular Edema Secondary to Retinal Vascular Disease: Preliminary Findings. IEEE Journal of Translational Engineering in Health and Medicine 2021;9.
13. Meng Z, Chen Y, Li H, et al. Machine learning and optical coherence tomography-derived radiomics analysis to predict persistent diabetic macular edema in patients undergoing anti-VEGF intravitreal therapy. Journal of Translational Medicine 2024;22:358.
14. Kar SS, Cetin H, Srivastava SK, et al. Optical coherence tomography-derived texture-based radiomics features identify eyes with intraocular inflammation in the HAWK clinical trial. Heliyon 2024;10(13).
15. Zarranz-Ventura J, Barraso M, Alé-Chilet A, et al. Evaluation of microvascular changes in the perifoveal vascular network using optical coherence tomography angiography (OCTA) in type I diabetes mellitus: a large scale prospective trial. BMC Med Imaging 2019;19:91.
16. Allan S, Olaiya R, Burhan R. Reviewing the use and quality of machine learning in developing clinical prediction models for cardiovascular disease. Postgraduate Medical Journal 2022;98:551–558.
17. Poplin, R., Varadarajan, A. v., Blumer, K., et al. Prediction of cardiovascular risk factors from retinal fundus photographs via deep learning. Nature Biomedical Engineering 2018;2(3):158–164.
18. Ting DSW and Wong TY. Eyeing cardiovascular risk factors. Nature Biomedical Engineering 2018;2(3):140–141.
19. Cheung CY, Xu D, Cheng CY, et al. A deep-learning system for the assessment of cardiovascular disease risk via the measurement of retinal-vessel calibre. Nature Biomedical Engineering 2021;5(6):498–508.
20. Tseng RMWW, Rim TH, Shantsila E, et al. Validation of a deep-learning-based retinal biomarker (Reti-CVD) in the prediction of cardiovascular disease: data from UK Biobank. BMC Medicine 2023;21(1).

21. Qian Y, Li L, Nakashima Y, et al. Is cardiovascular risk profiling from UK Biobank retinal images using explicit deep learning estimates of traditional risk factors equivalent to actual risk measurements? A prospective cohort study design. BMJ Open 2024;14(10):e078609.
22. Ghenciu LA, Dima, M, Stoicescu ER, et al. Retinal Imaging-Based Oculomics: Artificial Intelligence as a Tool in the Diagnosis of Cardiovascular and Metabolic Diseases. Biomedicines 2024;12(9).
23. Wu JH, Liu TYA. Application of deep learning to retinal-image-based oculomics for evaluation of systemic health: A review. Journal of Clinical Medicine 2023;12:152.
24. Huang S, Bacchi S, Chan W, et al. Detection of systemic cardiovascular illnesses and cardiometabolic risk factors with machine learning and optical coherence tomography angiography: a pilot study. Eye 2023;37:3629–3633.
25. Arnould L, Guenancia C, Bourredjem A, et al. Prediction of cardiovascular parameterswith supervised machine learning fromsingapore "i"vessel assessment and oct-angiography: A pilot study. Translational Vision Science & Technology 2021;10(13):20.
26. Barraso M, Alé-Chilet A, Hernández T, et al. Optical Coherence Tomography Angiography in Type 1 Diabetes Mellitus. Report 1: Diabetic Retinopathy. Translational Vision Science & Technology 2020;9(10):34.
27. Alé-Chilet A, Bernal-Morales C, Barraso M, et al. Optical Coherence Tomography Angiography in Type 1 Diabetes Mellitus—Report 2: Diabetic Kidney Disease. Journal of Clinical Medicine 2021;11(1):197.
28. Bernal-Morales C, Alé-Chilet A, Martín-Pinardel R, et al. Optical Coherence Tomography Angiography in Type 1 Diabetes Mellitus. Report 4: Glycated Haemoglobin. Diagnostics 2021;11(9):1537.
29. Hossain ME, Uddin S, Khan A. Network analytics and machine learning for predictive risk modelling of cardiovascular disease in patients with type 2 diabetes. Expert Systems with Applications 2021;164:113918.
30. Nabrdalik K, Kwiendacz H, Drozdz K, et al. Machine learning predicts cardiovascular events in patients with diabetes: The silesia diabetes-heart project. Expert Systems with Applications 2023;48(7):101694.
31. Hollmann N, Müller S, Purucker L, et al. Accurate predictions on small data with a tabular foundation model. Nature 2025;637:319-326.